# Double electron-positron pair production with two-electron capture in relativistic heavy-ion collisions

Melek YILMAZ ŞENGÜL

Atakent Mahallesi, 3.Etap, 34303 Halkalı-Küçükçekmece İstanbul,TURKEY

melekaurora@yahoo.com

We calculated the cross section of double electron-positron pair production with two-electron capture for the collisions of Pb+Pb ions and we did our calculations at LHC energies. We applied a similar methodology for the calculation of bound-free electron-positron pair production. We used perturbation theory and implemented Monte Carlo integration techniques to calculate the lowest order Feynman diagrams. We also compared our double electron-positron pair production with two-electron capture cross section results obtained in the literature. These calculations may help us to learn more about strong QED.

## 1 Introduction

In heavy ion colliders, heavy ions with a large charge ($Z_{x,y}$) create strong electromagnetic fields and because of this reason, electromagnetically produced lepton pairs have an enormous cross sections. If the negatively charged lepton is captured by one of the heavy ions, this would cause the depletion of the beam and the charge of the heavy ion would change and also it would not stay in the beam pipe. Several calculations were performed about the bound-free pair production [1-9] and the cross section can be defined as

$$\sigma_{BFPP} \propto Z_x^5 Z_y^2 \ln(\frac{\gamma}{\Delta}), \tag{1}$$

where $Z_x$, $Z_y$ and $\gamma$ represent the projectile atomic number, the target atomic number and the energy value, respectively, and $\Delta$ is a slowly varying parameter [10]. The electron capture into the projectile 1s shell with the positron emission process cross section varies as the square of the perturbing charge, as for most momentum transfer processes such as ionization and excitation, and as $Z_x^5$, which is the usual behaviour of electron capture processes for projectile charge [11].

Investigation of the multiple pair production is important. At high energy levels, the single pair production cross section becomes large and the contribution of producing more than one pair production probability is significant in the measured cross section. When the multiple pair production is calculated, then it can be understood if it has important contribution or not [12]. Also, multiple pair production processes may be the mixed form of the pair production process with electron capture and the number of free lepton pair production processes [13]. Investigation of the multiple pair production with BFPP will extend our knowledge of the non-linear QED effects in quantum vacuum. This process can be investigated at the LHC, because this heavy ion collider has forward detectors and they give us a chance of counting ions efficiently. As a result, these will help us to learn new information about the strong fields [1].

To give a help to the future LHC experiments, in this work we calculated double electron-positron pair production with two-electron capture for Pb+Pb collisions at LHC. Double electron-positron pair production with two-electron capture can be created in two ways. First, the electrons may be captured by the same ions as in (2)-(3)

$$Z_x + Z_y \rightarrow (Z_x + e^- + e^-)_{1s_{1/2}} + Z_y + e^+ + e^+ \qquad (2)$$

or

$$Z_x + Z_y \rightarrow Z_x + e^+ + e^+ + (Z_y + e^- + e^-)_{1s_{1/2}}. \qquad (3)$$

Second, the electrons may be captured by two different ions as in (4)

$$Z_x + Z_y \rightarrow (Z_x + e^-)_{1s_{1/2}} + (Z_y + e^-)_{1s_{1/2}} + e^+ + e^+. \qquad (4)$$

These three processes can be searched by detecting residual helium-like or two hydrogen-like ions experimentally and it is not necessary to observe the positrons [1, 14]. Also, in the future it will be possible to investigate these processes because of large cross sections that may reach to the 10 mb order [1].

**2 Formalism**

We worked on the double electron-positron pair production with two-electron capture in which electrons may be captured into the bound states of the same ions

$$Z_{x,y} + Z_{x,y} \rightarrow (Z_{x,y} + e^- + e^-)_{1s_{1/2}} + Z_{x,y} + e^+ + e^+. \qquad (5)$$

In order to calculate the double bound-free pair production cross sections, firstly, we expressed the probability of the single bound-free pair production (BFPP) as a function of impact parameter and then we calculated the cross section of double electron-positron pair production with two-electron capture for the collisions of Pb+Pb ions at LHC energies.

In this work, we did our calculations in the range of quantum electrodynamics (QED) by utilizing the lowest order perturbation theory. BFPP process including direct and crossed terms is characterized by Feynman diagrams in lowest QED order. In order to compute these terms, free positron is defined by the Sommerfeld-Maue wave function,

$$\psi_q^{(+)} = N_+ [e^{i\vec{q}\cdot\vec{r}} \mathrm{u}_{\sigma_q}^{(+)}], \qquad (6)$$

Here $N_+$ is a normalization constant that accounts for the distortion of the wave function that is acceptable for $Z\alpha \ll 1$, where $\alpha = e^2/\hbar c \cong 1/137$ is the fine structure constant, $v_+$ is the velocity of the positron in the rest frame of the ion, $\mathrm{u}_{\sigma_q}^{(+)}$ is the spinor structure of the outgoing positron. $N_+$ can be defined as

$$N_+ = e^{-\pi a_+/2}\Gamma(1+ia_+), \text{ where } a_+ = \frac{Ze^2}{v_+} \tag{7}$$

The captured electron is defined by the Darwin wave function,

$$\psi^{(-)}(\vec{r}) = \left(1 - \frac{i}{2m}\vec{\alpha}.\vec{\nabla}\right)u\frac{1}{\sqrt{\pi}}\left(\frac{Z}{a_H}\right)^{3/2}e^{-Zr/a_H}. \tag{8}$$

where u and $a_H = 1/e^2$ is the spinor part of the captured electron and the Bohr radius of atomic hydrogen, respectively. The collision of two heavy ions is taken into account in the "collider frame" with its total momentum is zero [15].

The cross section of BFPP for the second order perturbation theory can be expressed as

$$\sigma_{BFPP} = \int d^2b \sum_{q<0}\left|\left\langle\psi^{(-)}|S|\psi_q^{(+)}\right\rangle\right|^2 = \frac{|N_+|^2}{4\beta^2}\frac{1}{\pi}\left(\frac{Z}{a_H}\right)^3 \sum_{\sigma_q}\int\frac{d^3q d^2p_\perp}{(2\pi)^5}\left(\mathbf{A}^{(+)}(q:\vec{p}_\perp) + \mathbf{A}^{(-)}(q:\vec{q}_\perp - \vec{p}_\perp)\right)^2 \tag{9}$$

with

$$\mathbf{A}^{(+)}(q:\vec{p}_\perp) = F(-\vec{p}_\perp:\omega_x)F(\vec{p}_\perp - \vec{q}_\perp:\omega_y)\tau_q(\vec{p}_\perp:+\beta), \tag{10}$$

and

$$\mathbf{A}^{(-)}(q:\vec{q}_\perp - \vec{p}_\perp) = F(\vec{p}_\perp - \vec{q}_\perp:\omega_y)F(-\vec{p}_\perp:\omega_x)\tau_q(\vec{q}_\perp - \vec{p}_\perp:-\beta). \tag{11}$$

The explicit form of the scalar fields associated with ions "x" and "y" in momentum space can be written in terms of the corresponding frequencies as

$$F(-\vec{p}_\perp:\omega_x) = \frac{4\pi Ze}{\left(\frac{Z^2}{a_H^2} + \frac{\omega_x^2}{\gamma^2\beta^2} + \vec{p}_\perp^2\right)}, \tag{12}$$

where $\omega_x$ is the frequency for the heavy ion "x",

$$F(\vec{p}_\perp - \vec{q}_\perp:\omega_y) = \frac{4\pi Ze\gamma^2\beta^2}{\left(\omega_y^2 + \gamma^2\beta^2(\vec{p}_\perp - \vec{q}_\perp)^2\right)}, \tag{13}$$

where $\omega_y$ is the frequency for the heavy ion "y". For both ions "x" and "y", the transition amplitude can be represented as

$$\tau_q(\vec{p}_\perp;+\beta) = \sum_s \sum_{\sigma_p} \frac{1}{\left(E_p^{(s)} - \left(\frac{E^{(-)} + E_q^{(+)}}{2}\right) - \beta\frac{q_z}{2}\right)} \left[1 + \frac{\vec{\alpha}.\vec{p}}{2m}\right] \tag{14}$$

$$< u|(1-\beta\alpha_z)|u_{\sigma_p}^{(s)} >< u_{\sigma_p}^{(s)}|(1+\beta\alpha_z)|u_{\sigma_q}^{(+)} > .$$

This term relates the intermediate photon lines to the outgoing electron-positron lines. Transition amplitude depends explicitly on the velocity of the ions $(\beta)$, the transverse momentum of the intermediate-state $(\vec{p}_\perp)$, the parallel momentum of the intermediate-state $(p_z)$ and the momentum of the positron $(q)$. $u_{\sigma_p}^{(s)}$ is the spinor part of the intermediate-state [15].

In our previous calculations [10], we derived the cross section as a function of impact parameter for BFPP as seen below

$$\frac{d\sigma_{BFPP}}{db} = \int_0^\infty dq\, qb\, J_0(qb) F(q). \tag{15}$$

This equation includes a highly oscillatory Bessel function of order zero and the function $F(q)$ is a six-dimensional integral as written below

$$F(q) = \frac{\pi}{8\beta^2}|N_+|^2 \frac{1}{\pi}\left(\frac{Z}{a_H}\right)^3 \sum_{\sigma_q} \int_0^{2\pi} d\phi_q \int \frac{dq_z d^2K d^2Q}{(2\pi)^7}$$

$$\times \left\{ \begin{array}{l} F\left[\frac{1}{2}(\vec{Q}-\vec{q});\omega_x\right] F\left[-\vec{K};\omega_y\right] \tau_q\left[-\frac{1}{2}(\vec{Q}-\vec{q});+\beta\right] \\ + F\left[\frac{1}{2}(\vec{Q}-\vec{q});\omega_x\right] F\left[-\vec{K};\omega_y\right] \tau_q[\vec{K};-\beta] \end{array} \right\}$$

$$\times \left\{ \begin{array}{l} F\left[\frac{1}{2}(\vec{Q}+\vec{q});\omega_x\right] F\left[-\vec{K};\omega_y\right] \tau_q\left[-\frac{1}{2}(\vec{Q}+\vec{q});+\beta\right] \\ + F\left[\frac{1}{2}(\vec{Q}+\vec{q});\omega_x\right] F\left[-\vec{K};\omega_y\right] \tau_q[\vec{K};-\beta] \end{array} \right\} \tag{16}$$

For these calculations, $\vec{Q}$ and $\vec{K}$ are the new variables that are the functions of $\vec{q}$ and $\vec{p}$. After integrating (16) numerically, a very simple relation for $F(q)$ is obtained for a fixed value of $q$ as expressed below

$$F(q) = F(0) e^{-aq} = \sigma_{BFPP}\, e^{-aq}. \tag{17}$$

In the above equation, $F(0)$ is the value of the function at $q=0$ and equals to the total cross section of BFPP. The slope of the function $F(q)$ is a constant which is independent from the charges and energies of the heavy ions and as a good approximation it is equal to $a = 1.35\lambda_C$ [16]. The probability as a function of impact parameter for BFPP process can be written as,

$$P_{BFPP}(b) = \frac{1}{2\pi b} \frac{d\sigma}{db} = \frac{1}{2\pi} C_\infty \lambda_C^2 Z_x^2 Z_y^2 \alpha^4 \ln^3(\gamma) \frac{a}{(a^2+b^2)^{3/2}} = \sigma_{BFPP} \frac{a}{2\pi(a^2+b^2)^{3/2}}. \quad (18)$$

Here $C_\infty = 2.19$ is the fitted parameter, $\sigma_0 = \lambda_C^2 Z_x^2 Z_y^2 \alpha^4$ is the reduced cross section, where $\lambda_C = \hbar/mc$ is the reduced Compton wavelength of the electron, $Z_x$ and $Z_y$ are the charges of the colliding ions. $\gamma - 1$ (for all $\gamma > 3$) is the beam kinetic energy per nucleon in units where the nucleon mass is 1 [16].

In our calculations, $P_{BFPP}(b)$ and $P_{2ee}(b)$ expresses one and two BFPP probability, respectively. Two scenarios can be tought in these collisions. In the first scenario, both free electrons are captured by the same heavy ion, as in (2) and (3). The probability of this process can be shown as the product of the single bound free pair production probabilities. This can be expressed as

$$P_{2ee}(b;Z_y,Z_x) = \frac{1}{2} P_{BFPP}(b;Z_y,Z_x) P_{BFPP}(b;Z_y,Z_x) = \frac{1}{2} \frac{\sigma_{BFPP}\, a}{2\pi} \frac{1}{(a^2+b^2)^{3/2}} \frac{\sigma_{BFPP}\, a}{2\pi} \frac{1}{(a^2+b^2)^{3/2}} \quad (19)$$

and the cross section can be written as

$$\sigma_{2ee} = \int_0^\infty P_{2ee}(b)\, d^2b = \frac{1}{2}\int_0^\infty P_{BFPP}(b) P_{BFPP}(b)\, 2\pi b\, db, \quad (20)$$

also the detailed calulations of all probabilities and cross sections can be found in [1].

In the second scenario, free electrons can be captured by each heavy ion that causes the formation of the two hydrogen-like ions as in (4) and the probability can be written simply as

$$P_{ee+ee}(b;Z_x,Z_y) \approx P_{BFPP}(b;Z_y,Z_x) P_{BFPP}(b;Z_x,Z_y) \quad (21)$$

and the total cross section given in (4) can be expressed as

$$\sigma_{ee+ee} = 2\sigma_{2ee} = \int_0^\infty |P_{BFPP}(b)|^2\, 2\pi b\, db. \quad (22)$$

As discussed and proved detaily in [1], the cross section results for helium-like atom are related to the cross section results for two hydrogen-like atoms, and their relationship are given in (22).

**3 Results and Discussions**

For the calculation of the double lepton pair production cross section, we worked on the range of QED perturbation theory. We calculated the lowest order Feynman diagrams by using Darwin wave functions for the captured electron and Sommerfeld-Maue wave functions for the free positron. $F(q)$ function that is mentioned above is calculated by using Monte Carlo techniques and the integrals are tested on about 10 M randomly chosen "positions" in order to converge to our theoretical results. The numerical errors in these calculations are approximately five percent or less. Then, we expressed the single pair production probability $P_{BFPP}$ as a function of impact parameter for BFPP process. By using this expression, we calculated the cross section of double electron-positron pair production with two-electron capture process as expressed in (22). Finally, we compared our double electron-positron pair production with two-electron capture process cross section results for the collisions of Pb+Pb ions to [1]. The compared results are in agreement with each other. In [1], the total cross section of two processses using the relativistic partial-wave calculations is found upon numerical integration over the impact parameter as $\sigma_{ee+ee} = 2\sigma_{2ee} = 11\,mb$. The same processes are calculated by using the equivalent photon approximation. As mentioned in [1], the large contribution to this integral comes from the small impact parameter region as expressed in [1]. The cross section result is equal to $\sigma_{ee+ee} = 2\sigma_{2ee} = 12,6\,mb$. We did the same calculations by using the lowest order QED. To calculate cross section results, we used Monte Carlo method. Our result is $\sigma_{ee+ee-Ourwork} = 2\sigma_{2ee-Ourwork} = 11,57\,mb$. When we compare our results with the other work done in [1], it is clear that they are very close to each other. The discrepancy between these two results is approximately $5\%$. While taking the integral to reach the cross section results, we also proved that the main contrubution to this integral comes from the small impact parameter region as shown in Fig. 1. In Fig.1, we plotted the impact parameter dependence probability function by using (18). When we compare our probability function behaviour (as seen in Fig.1) with the impact parameter dependent pair production probability behaviour which is given in [1], it is seen that our probability function goes to zero more quickly than the function in [1]. The reason for this difference can be explained as follows: For the large impact parameters, our probability function behaves like $\approx 1/b^3$, but the other one given in [1] behaves as $\approx 1/b^2$.

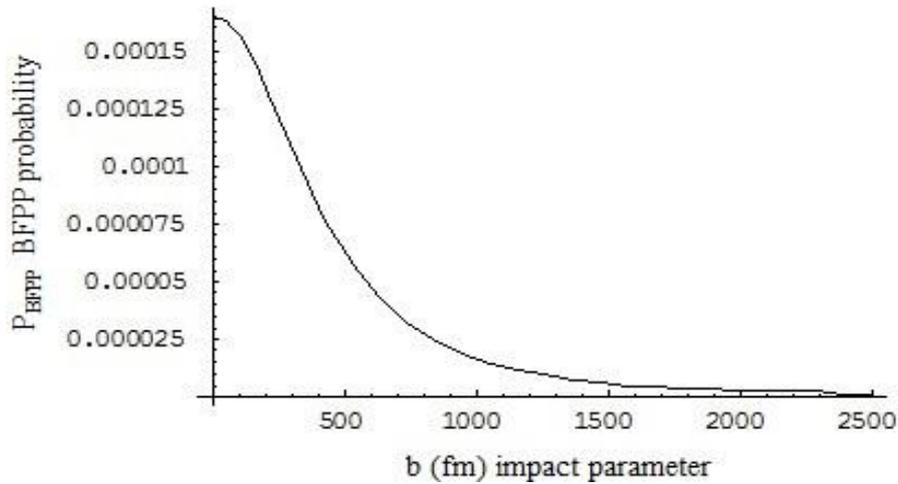

**Fig.1.** The probability $P_{BFPP}(b)$ of a bound-free pair production in the collisions of Pb+Pb.

**4 Conclusions**

In [1], the cross section of double electron-positron pair production with two-electron capture has been calculated by using two different methods. In the first method, the cross section calculation is done by using the first order perturbation theory and the partial-wave expansion of the Dirac wave functions. In the second method, for the cross section calculation an analytical expression is derived in the range of the equivalent photon approximation. Calculations are done for the collisions of the Pb+Pb ions for the LHC. The probability expressions have been derived for a single helium-like ions in (2) and (3) or two hydrogen like ions in (4). In this work, we studied on the electron-positron pair production with the capture of the electron into the K shell of one of the ions. In the framework of

QED perturbation theory, we calculated the lowest order Feynman diagrams. We obtained an analytical expression for the impact parameter dependent BFPP probability. The importance of this work can be summarized by expressing the impact parameter dependent probability of double electron-positron pair production with two-electron capture, and by calculating the cross section of this process for the collisions of the Pb+Pb ions at the LHC energy numerically by using Monte Carlo Methods. Also, two methods described above and explained detaily in [1] are in a good agreement with our method given in this work. These cross section calculations will help us to understand the new information on the QED in strong electromagnetic fields.

## REFERENCES

**[1]** A.N. Artemyev, V.G. Serbo, and A. Surzhykov, Eur. Phys. J. C,72, (2012), arXiv:1204.0263.

**[2]** C.A. Bertulani, and D.Dolci, Nucl. Phys. A, 683, 635, (2001).

**[3]** A. J. Baltz, M. J. Rhoades-Brown, and J. Weneser, Phys. Rev. A , 50, 4842, (1994).

**[4]** M. J. Rhoades-Brown., C. Bottcher, and M.R. Strayer, Phys. Rev. A , 40, 2831, (1989).

**[5]** H. Meier, Z. Halabuka, K. Hencken, D. Trautmann, and G. Baur, Phys. Rev. A , 63, 032713, (2001).

**[6]** R. Bruce, J. M. Jowett, S. Gilardoni, A. Drees, W. Fischer, S.Tepikian, and S.R.Klein, Phys. Rev. Lett., 99, 144801, (2007).

**[7]** S.R.Klein, Nucl. Inst. and Meth. in Phys. Res. A, 459, 51, (2001).

**[8]** A. Belkacem, H. Gould, B. Feinberg, R. Bossingham, and W.E. Meyerhof, Phys. Rev. Lett., 71, 1514, (1993).

**[9]** H.F. Krause, C.R. Vane, S. Datz, P. Grafström, H. Knudsen, C. Scheidenberger, and R.H.Schuch, Phys. Rev. Lett., 80, 1190, (1998).

**[10]** M.Y. Şengül, and M. C. Güçlü, Phys. Rev. C, 83, 014902, (2011).

**[11]** R. Anholt, and U. Becker, Phys. Rev. A , 36, 4628 (1987).

**[12]** M. C. Güçlü , J. Li, A.S. Umar, D.J. Ernst, and M.R. Strayer, Annals of Phys., 272, 7, (1999).

**[13]** G. Baur, K. Hencken, D. Trautmann, S. Sadovsky, and Y. Kharlov, Phys. Rept., 364, 359 (2002).

**[14]** R. Schicker, arXiv:1512.02060v1.

**[15]** M.Y. Şengül, M.C. Güçlü, and S. Fritzsche, Phys. Rev. A, 80, 042711, (2009).

**[16]** M. C. Güçlü, Nucl. Phys. A, 668,149, (2000).